\documentclass[aps,prd,preprint,groupedaddress,longbibliography]{revtex4-1}
\usepackage{amsfonts,amsmath,amssymb}
\usepackage{graphicx}

\newcommand{\Lagr}{\mathcal{L}}
\def\exd{\mathrm{d}}

\begin{document}

%\preprint{}

\title{On the thermodynamics of universal horizons in Einstein-{\AE}ther theory}


\author{Arif Mohd}
\email[]{arif.mohd@sissa.it}
%\thanks{}
\affiliation{SISSA - International School for Advanced Studies, \\
                  Via Bonomea 265, 34136 Trieste, Italy \\
                                       and  \\
                INFN, Sezione di Trieste, Trieste, Italy.}


\date{\today}

\begin{abstract}
The theories of gravity which violate local Lorentz invariance do not admit a universal maximum speed of signal-propagation. Different field excitations see a different effective metric and hence a different light cone. In these theories,  although one can define the Killing horizon in a conventional way, this definition does not capture the notion of a black hole. This is so because there exist modes which see a wider light cone than the one defined by the Killing Horizon and therefore can escape to infinity. However, there exist solutions of these theories which admit a special spacelike hypersurface which acts as a one-way membrane. Signals from beyond this hypersurface can never escape to infinity and are destined to hit the singularity. In this sense this hypersurface acts like a black-hole horizon and is called the Universal Horizon because it traps modes travelling with arbitrarily high velocities. We use the Noether charge method {\it \`a la} Wald to show that a first law, which resembles the first law of thermodynamics, can be formulated for universal horizons in the Einstein-{\AE}ther theory. This seems to suggest that in Lorentz violating theories one should ascribe the thermodynamical properties to the universal horizon and not to the Killing horizon.
\end{abstract}

%\keywords{}

\maketitle

\section{Introduction}
Invariance under Lorentz transformations is a fundamental symmetry of the quantum field theories describing the elementary-particle interactions in nature. There are very strict observational and experimental constraints on the violation of Lorentz invariance \cite{lrr-2005-5, Eling:2004dk}. However, it still leaves open the possibility that the Lorentz symmetry is fundamental but the vacuum does not respect it. In other words, some field(s) of the underlying Lorentz invariant theory could acquire vacuum expectation value(s) and thus the Lorentz symmetry would be spontaneously broken \cite{Kostelecky:1988zi}. Such a scenario is specially attractive when gravity is included because it opens the door to the renormalizability of perturbative quantum gravity, at least at the level of power counting \cite{Horava:2009uw}. \par
Einstein-{\AE}ther theory is a diffeomorphism invariant theory of gravity which violates the local Lorentz invariance by introducing a dynamical vector field $u^a$ which is constrained to be unit-timelike everywhere \cite{Jacobson:2000xp}. This vector field is called the {\ae}ther. The theory respects general covariance because the {\ae}ther is dynamical. The theory violates the local Lorentz invariance because the {\ae}ther introduces a preferred frame (albeit dynamically determined) by virtue of the constraint that it is unit-timelike everywhere. It is an effective field theory which views the {\ae}ther as representing some kind of a condensate that spontaneously breaks the local Lorentz invariance.  The observational constraints on the Einstein-{\AE}ther theory are discussed in ref.~\cite{Jacobson:2008aj}. \par
The appearance of a preferred frame has interesting consequences. Now it is allowed for the matter to couple with the {\ae}ther which leads to the modified dispersion relation. This in turn leads to the superluminal propagation of particles. One can then do the gedanken experiments and create the perpetual-motion machines operating either quantum mechanically via the Hawking's process \cite{Dubovsky:2006vk} or classically using a Penrose-like process \cite{Eling:2007qd,Jacobson:2008yc} thus leading to a violation of the Genearlized Second Law (GSL) \cite{Bekenstein:1974ax}. \par
Lorentz invariance and causality are of course completely different things \cite{Coleman:1969xz}. Superluminal propagation does not mean that the theory violates causality. For causality, one needs a definition of light-cone. In the Lorentz invariant theories it is natural to use the invariant speed of light (c) to define the light cones. The signals which propagate inside these light cones are considered to be causal while the others are deemed acausal. In Lorentz violating theories, like the Einstein-{\AE}ther theory, there is no maximal speed that can be used to define a universal ``light-cone".  However, since one has a preferred frame at disposal one uses that to impose causality. In particular, the signals which propagate towards the future of the local hypersurface orthogonal to the {\ae}ther are declared to be causal. This has a direct consequence that the notion of a black hole also needs to be modified. In Lorentz invariant theories the Killing horizon provides a good notion of the event horizon  which acts as a one-way membrane separating the inside and outside of a stationary black hole. If we have modes propagating with arbitrarily high speed in the theory, the Killing horizon is clearly no more a useful concept to define the black hole region in spacetime. \par
\begin{figure}
  \includegraphics[scale=1]{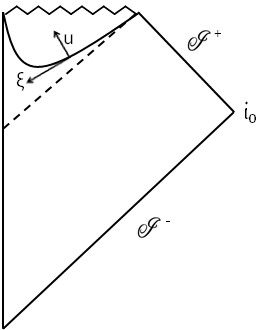}
  \caption{Universal Horizon is the hypersurface where the time translation Killing vector field $\xi^a$ becomes orthogonal to the {\ae}ther $u^a$.} 
\label{fig:UnivHorizon}
\end{figure}

Consider, however, the static and spherically symmetric spacetimes in which the {\ae}ther is hypersurface orthogonal \cite{Eling:2006ec}. At infinity, the {\ae}ther and the time-translation Killing vector $\xi$ are aligned. Inside the Killing horizon $\xi$ becomes spacelike. Consider that particular hypersurface where  $\xi$ becomes orthogonal to the {\ae}ther and hence is tangent to this hypersurface normal to the {\ae}ther (see fig.~\ref{fig:UnivHorizon}). Any causal signal (i.e., one which propagates in the future of this hypersurface) necessarily moves towards a decreasing radius and eventually hits the singularity. This hypersurface acts as a causal boundary which acts like a one-way membrane. Any signal, moving at arbitrarily high speed, after crossing it can never escape to infinity. This special hypersurface where the time-translational Killing field becomes orthogonal to the {\ae}ther is called the Universal Horizon. Since it traps the modes of arbitrarily high velocities, the universal horizon defines a causal boundary and hence the black hole region in spacetime. A regular universal horizon is seen to exist in the one-parameter family of static spherically symmetric solutions of the Einstein-{\AE}ther theory \cite{Eling:2006ec,Barausse:2011pu}. Universal horizons in Ho\v{r}ava gravity are discussed in ref.~\cite{Blas:2011ni}. \par
The presence of modified dispersion relations and universal horizons in the Einstein-{\AE}ther theory (and the H\v{o}rava gravity) also provides a possibility to evade the arguments creating the perpetual-motion machines. These arguments involve the finite maximal speed (different than c) for different particles and the thermodynamics associated to the effective black-hole geometry that these particles see. If, however, it turns out that due to the modified dispersion relations there is no universal upper limit on the propagation speed of particles and that the unambiguous thermodynamics can be ascribed only to the universal horizons then one might hope to save the GSL. It was shown in ref.~\cite{Berglund:2012bu} that the universal horizons in static spherically symmetric solutions of the Einstein-{\AE}ther theory follow a first-law kind of relation. The role of entropy is played by a quarter of the area of the horizon. In ref.~\cite{Berglund:2012fk}, using the tunnelling method of ref.~\cite{Parikh:1999mf}, it was shown that the quantity playing the role of the temperature in the first law is really the temperature at which the universal horizon radiates. This already provides a strong hint that in the Lorentz violating theories the laws of black hole thermodynamics can be ascribed to the universal horizons.
\par
The goal of this paper is to use the Noether charge method of Wald to derive the first-law of mechanics for the universal horizons in the static, spherically symmetric solutions of the Einstein-{\AE}ther theory. The first law was first proved in ref.~\cite{Berglund:2012bu}, using the methods akin to ref.~\cite{Bardeen:1973gs}, by manipulating the various projections of the equations of motion of the theory. On the other hand, we have an elegant geometric Noether-charge method of Wald to derive the first law for any diffeomorphism-invariant theory of gravity admitting a regular bifurcation surface. Herein lies the difficulty because in the Einstein-{\AE}ther theory the bifurcation surface is not regular: the {\ae}ther necessarily diverges there \cite{Foster:2005fr}. Indeed, the Noether-charge method was used in ref.~\cite{Foster:2005fr} to prove a first law for the Killing horizons in the Einstein-{\AE}ther theory but no thermodynamic interpretation emerged in that study. We will show that we can nevertheless use the Noether-charge method to prove a first law for the universal horizons which, according to ref.~\cite{Berglund:2012fk}, has a natural thermodynamic interpretation.\par
 This paper is organized as follows: in sec.~\ref{sec:Wald's formalism} we summarize the Wald's formalism and we review why it does not apply to the universal horizons in a straight-forward fashion. In sec.~\ref{sec:EA theory} we construct the Noether charge conjugate to the diffeomorphisms in the Einstein-{\AE}ther theory. In sec.~\ref{sec:first law} we prove the first-law for universal horizons in the static, spherically- symmetric solutions. We conclude with a summary and open questions in sec.~\ref{sec:Discussion}. The consequences of staticity and spherical symmetry are discussed in the appendix and the equations given there are used repeatedly in sec.~\ref{sec:first law}.
\section{Wald's formalism}
\label{sec:Wald's formalism}
In this section we review Wald's formalism for constructing the Noether charge corresponding to the diffeomorphism invariance of the theory. We will be viewing the integrands as appropriate tensor densities \cite{1990JMP....31.2378W}. In particular, 
Lagrangian is a scalar density of weight one \footnote{We remind the reader that the Lie-derivative of a vector density $T^a$ of weight $w$ is given by
$\mathcal \pounds_\xi T^a = \xi^c \nabla_c T^a - T^c \nabla_c \xi^a + w\, T^a \nabla_c \xi^c.$ }. 
The dynamical fields of the theory are denoted by $\phi^i$. \par
 A general variation of the Lagrangian density gives
\begin{align}
\delta \Lagr = E_i \delta \phi^i + \nabla_a \theta^a.
\end{align}
The equations of motion are given by $E_i=0$. The surface term $\theta^a$ is called the symplectic potential current density.
Due to the commutativity of two variations, $\delta_1 \delta_2=\delta_2 \delta_1$, we have
\begin{align}
(\delta_1 \delta_2 - \delta_2\delta_1) \Lagr &= \delta_1 E_i \delta_2 \phi^i - \delta_2 E_i \delta_1 \phi^i + \nabla_a \omega^a \nonumber \\
 &= 0, 
\end{align} 
where $\omega^a= \delta_1 \theta^a (\delta_2) - \delta_2 \theta^a (\delta_1) $ is the symplectic current density.
 When the dynamical field satisfies the equation of motion, i.e., $E_i = 0$ and the linearized equation of motion is also satisfied, i.e., $\delta_1 E_i = 0 = \delta_2 E_i$ then the symplectic current density is conserved, $\nabla_a \omega^a = 0$. Integral of $\omega^a$ over a Cauchy surface $\Sigma$ is called the symplectic current,
\begin{align}
\omega = \int_\Sigma \exd \Sigma_a \, \omega^a.
\end{align}
Diffeomorphism invariance of the theory means that the variation of the Lagrangian density under an infinitesimal diffeomorphism generated by a vector field $\xi$ is given by the Lie derivative of the Lagrangian density along $\xi$,
\begin{align}
\delta_\xi \Lagr = \nabla_a(\xi^a \Lagr).
\end{align}
Now associate a Noether-current density $J_\xi^a$ corresponding to the variation induced by the diffeomorphism as 
\begin{align}
J_\xi^a = \theta^a(\delta_\xi) - \xi^a \Lagr.
\end{align}
Then it can be checked that the Noether-current density is conserved on shell,
\begin{align}
\nabla_a J_\xi ^a & = \nabla_a \theta^a(\delta_\xi) - \nabla_a(\xi^a \Lagr) \nonumber \\
& = \delta_\xi \Lagr + E_i \delta_\xi \phi^i - \nabla_a(\xi^a \Lagr) \nonumber \\
& = E_i \delta_\xi \phi^i \nonumber \\
& = 0 \,\,\,\,\,\, \,\,\,\text{on-shell}.
\end{align}
This implies that there exists an antisymmetric tensor density $Q_\xi^{ab}$ which acts as a potential for the Noether charge, in the sense that when the dynamical fields satisfy the equations of motion then
\begin{align}
J_\xi^a = 2\, \nabla_b Q_\xi^{ab}.
\end{align}
The integral of $J_\xi ^a $ over a Cauchy surface defines the corresponding Noether charge,
\begin{align}
Q_\xi = \int_\Sigma \exd \Sigma_a \, J_\xi^a.
\end{align}
When the equations of motion are satisfied we can replace the integrand by the Noether potential and use the Stoke's theorem to get,
\begin{align}
Q_\xi = \int_{\partial \Sigma} \exd \sigma_{ab} \, Q^{ab}_\xi. 
\end{align}
Now consider an arbitrary variation of $J_\xi^a$,
\begin{align}
\delta J_\xi^a = \delta \theta^a(\delta_\xi) - \delta(\xi^a \cal{L}), \nonumber
\end{align}
where the variation $\delta$ acts only on the dynamical fields of the theory, in particular $\delta \xi^a =0$. This gives
\begin{align}
\delta J_\xi^a = \omega^a(\delta, \delta_\xi) + 2 \,\nabla_b (\theta^{[a} \xi^{b]} ).
\end{align}
If the Hamiltonian corresponding to the evolution by $\xi$ exists on the phase space then by its very definition 
\begin{align}
\label{eq:hamiltonian definition}
\delta H_\xi = \int_\Sigma \exd \Sigma_a \, \omega^a(\delta, \delta_\xi),
\end{align}
we get, 
\begin{align}
\delta H_\xi =  \int_\Sigma \exd \Sigma_a \left( \delta J_\xi^a -2 \,\nabla_b (\theta^{[a}\xi^{b]}) \right).
\end{align}
On shell this expression reduces to a surface term,
\begin{align}
\label{eq:hamiltonian_wald}
\delta H_\xi = \int_{\partial \Sigma} \exd \sigma_{ab}\, \left( \delta Q_\xi^{ab} - \,\theta^{[a}\xi^{b]}\right).
\end{align}
If one can find a quantity $B^a$ such that $\int_{\partial \Sigma}\exd \sigma_{ab}\,\theta^{[a}\xi^{b]} = \delta \left( \int_{\partial \Sigma}\exd \sigma_{ab}\,B^{[a}\xi^{b]} \right)$ then one can identify the Hamiltonian generating the motion in the phase space induced by $\xi$ as
\begin{align}
H_\xi = \int_{\partial \Sigma} \exd \sigma_{ab}\,\left( Q_\xi^{ab} -\,B^{[a}\xi^{b]} \right).
\end{align}
If $\xi$ is an asymptotic  time translation in an asymptotically flat space then the corresponding Hamiltonian is identified as the canonical energy $\mathcal{E}$ of the spacetime. If $\xi$ is an asymptotic rotation then the corresponding Hamiltonian is identified as the negative of the angular momentum $\mathcal{J}$ of the spacetime. \par
\begin{figure}
\includegraphics[scale=0.75]{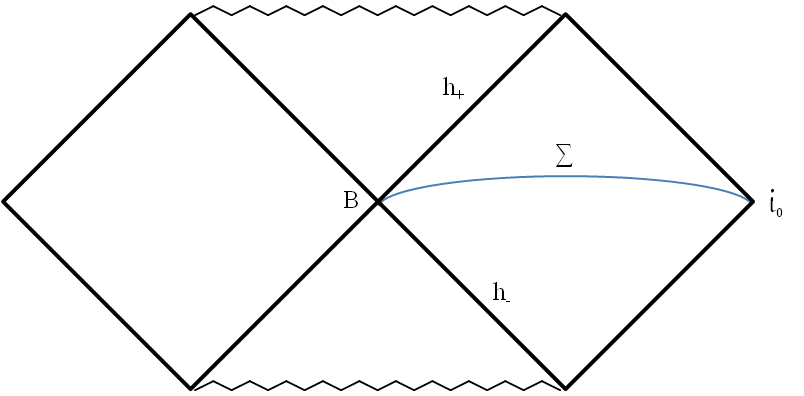}
\caption{A slice $\Sigma$ extending from the spacelike infinity $i_o$ to the bifurcation surface $B$.}
\label{fig:CauchySlice}
\end{figure}

Now consider a stationary black-hole solution with a bifurcate Killing horizon and the bifurcation surface $S_B$. Let $t^a$ be the stationary Killing vector field and $\varphi^a$ be the axial Killing vector field.  Let $\xi^a=t^a + \Omega \varphi^a$  be such that it vanishes on the bifurcation surface. $\Omega$ is the angular velocity of the horizon. If all the fields in the theory are Lie dragged by $\xi^a$ then eq.~\eqref{eq:hamiltonian definition} implies that $\delta H_\xi = 0$. Now choose $\Sigma$ to be any spacelike hypersurface that runs from the asymptotic spatial infinity to the bifurcation surface $B$ (see fig.~\ref{fig:CauchySlice}). Then noting that $\xi^a$  vanishes on $B$, eq.~\eqref{eq:hamiltonian_wald} implies that
\begin{align}
\label{eq:first law}
\delta \oint_B \exd \sigma_{ab}\,   Q_\xi^{ab} &= \delta \oint_{\infty} \exd \sigma_{ab}\, \left(Q_\xi^{ab} - \,B^{[a}\xi^{b]}\right) \nonumber \\
&= \delta \oint_{\infty} \exd \sigma_{ab}\, \left(Q_t^{ab} - \,B^{[a}t^{b]}\right) +\delta \oint_{\infty} \exd \sigma_{ab}\, Q_\varphi^{ab},
\end{align}
where the integrals are on the boundary of the slice $\Sigma$: one is the bifurcation surface $B$ and the other is the sphere at the asymptotic infinity. Now recognizing the two terms on the right hand side as $\delta \mathcal{E}$ and $-\delta \mathcal{J}$ where $\mathcal{E}$ and $\mathcal{J}$ are the asymptotically defined Energy and Angular Momentum, respectively. This almost resembles the first law of black-hole mechanics. However, the left-hand side is not of the form of $\kappa$ times the variation of a local, geometric quantity on $B$. There is further work required to write it in that form, but for our purposes we are not going to need it here. Assuming that one can write the left hand side as $\kappa/2\pi$ times the variation of an integral then the above equation is the first law of black hole mechanics
\begin{align}
\delta \mathcal{E} = \frac{\kappa}{2\pi} \delta \mathcal{S} + \Omega\mathcal{J},
\end{align}
where $\mathcal{S}$ is now interpreted as the entropy of the black hole in the theory. \par
In order for Wald's procedure to go through it is important that all the fields in the theory are Lie-dragged by the Killing vector field $\xi^a$ and that all the fields admit a regular extension to the bifurcation surface. The later requirement can not be met by the {\ae}ther field \cite{Foster:2005fr}. The reason is that on the bifurcation surface the Lie-drag acts like a radial boost. Thus the only vectors on the bifurcation surface that can be Lie-dragged are the ones tangent to the bifurcation surface which are spacelike. Since the {\ae}ther is constrained to be timelike it cannot be Lie-dragged. If we impose that it be Lie-dragged then it cannot be regular. This is the main difficulty in using Wald's method to prove the first law for the Killing horizons in the Einstein-{\AE}ther theory. But we can still use the Wald's method to look for a first law for the universal horizon and this is what we turn to next. We start by deriving an expression for the Noether charge conjugate to diffeomorphisms in the Einstein-{\AE}ther theory.

\section{Einstein-{\AE}ther theory}
\label{sec:EA theory}
Action for the Einstein-{\AE}ther theory is given by
\begin{align}
S= \frac{1}{16 \pi G_{\ae}}\int \exd^4x  \sqrt{-g}\, \left( R + L_{\ae} \right),
\end{align}
where the {\ae}ther-dependent part is 
\begin{align}
L_{\ae} = - {Z^{ab}}_{cd} \, \nabla_a u^c \, \nabla_bu^d + \lambda (u^2 +1).
\end{align}
Here $\lambda$ is  a Lagrangian multiplier that enforces the unit timelike normalization of the {\ae}ther four-vector $u^a$, and ${Z^{ab}}_{cd}$ describes the coupling of the {\ae}ther with the metric in terms of the coupling constants $c_i, i=1,2,3,4$, as
\begin{align}
{Z^{ab}}_{cd} = c_1\, g^{ab} g_{cd} + c_2\, \delta^a_c \delta^b_d + c_3 \,\delta^a_d  \delta^b_c - c_4\, u^a u^b g_{cd}.
\end{align} 
The weak-field limit \cite{Carroll:2004ai} can be used to relate constant $G_{\ae}$ occurring in the action and  the Newton's constant G as
\begin{align}
G_{\ae}= \left( 1 - \frac{c_{14}}{2} \right) G.
\end{align}
Einstein-{\AE}ther theory is a viable theory of gravity for a particular range of the coupling constants which we mention though we are not going to use it.  In ref.~\cite{Jacobson:2004ts}, by linearizing the theory on a flat background with a constant
{\ae}ther, and demanding that the squared speeds of the modes be positive, the range of the coupling constants was found to be
\begin{align}
0 \leq c_{13} < 1, \nonumber \\
0 \leq c_{14} < 2, \nonumber \\
2+c_{13}+3c_2>0.
\end{align} 
\par
Variation of action with respect to the metric gives the equation of motion for the metric,
\begin{align}
G^{ab} = T^{ab}_{\ae},
\end{align}
where $G^{ab} = R^{ab} - \dfrac{1}{2} R g^{ab}$ is the Einstein tensor and the stress tensor of the {\ae}ther is given by
\begin{align}
 T^{ab}_{\ae} = \frac{1}{2} L_{\ae} g^{ab} &+ c_1 (\nabla^a u_c \nabla^b u^c - \nabla_c u^a \nabla^c u^b) \\ \nonumber
&+ c_4 a^a a^b + \lambda u^a u^b + \nabla_c \left(Y^{c(b}u^{a)} + Y^{(ab)} u^c -u^{(b}Y^{a)c}\right),
\end{align}
where $a^b=u^c \nabla_c u^b$ is the acceleration of the {\ae}ther and ${Y^a}_c = {Z^{ab}}_{cd} \nabla_b u^d$. Variation of the action with respect to $u^a$ gives the equation of motion for the {\ae}ther,
\begin{align}
\lambda u_a + c_4 a^c \nabla_a u_c + \nabla_c {Y^c}_a = 0.
\end{align}
Collecting all the surface terms in the variation of the action gives the symplectic potential current density,
\begin{align}
\label{eq:theta_wald}
\theta^c = \frac{1}{16 \pi G_{\ae}} \sqrt{-g} &\, \Big[ g^{ca} \nabla^b(\delta g_{ab}) - g^{ab}\nabla^c (\delta g_{ab})  
\nonumber \\
                                                            & - \left(Y^{c(b}u^{a)} + Y^{(ab)} u^c -u^{(b}Y^{a)c}\right) \delta g_{ab} \nonumber \\
                                                            & - 2 {Y^c}_a \delta u^a \Big]. 
\end{align}
Noether current density corresponding to the infinitesimal diffeomorphism generated by a vector field $\xi$ can now be calculated. On shell this can be written as 
\begin{align}
J_\xi^c = 2 \, \nabla_a Q_\xi^{ca},
\end{align}
where $Q_\xi^{ca}$ is given by
\begin{align}
\label{eq:Noether expression for ae}
Q_\xi^{ca} = \frac{1}{16 \pi G_{\ae}} \sqrt{-g} \Big[- \nabla^{[c} \xi^{a]} - \xi_b \left(Y^{[ca]} u^b + u^{[c} Y^{a]b}-Y^{b[c} u^{a]}\right)\Big]
\end{align}
We saw at the end of sec.~\ref{sec:Wald's formalism} that in the Einstein-{\AE}theory the bifurcation surface is not a regular surface. So in the next section, we turn to proving the first-law for the universal horizon.
\section{First law for static, spherically-symmetric black holes}
\label{sec:first law}
\begin{figure}
\includegraphics[]{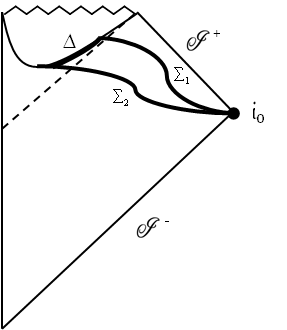}
\caption{A portion $\Delta$ of the Universal horizon sandwiched between two slices $\Sigma_1$ and $\Sigma_2$. }
\label{fig:Slice_UH}
\end{figure}
Consider a portion of spacetime $\mathcal{M}$ sandwiched between a portion of universal horizon $\Delta$ and two spacelike slices $\Sigma_1$ and $\Sigma_2$ (see fig.~\ref{fig:Slice_UH}). Since the symplectic structure is a closed form (by construction), i.e., the divergence $\nabla_a \omega^a$ vanishes, we have
\begin{align}
\int_{\mathcal{M}} \nabla_a \omega^a = 0, \nonumber \\
=> \int_{\Sigma_1} \exd \Sigma_a \, \omega^a - \int_{\Sigma_2} \exd \Sigma_a \, \omega^a + \int_\Delta \exd \Sigma_a \, \omega^a &= 0.
\end{align}
For general field variations the last integral does not vanish. But $\omega^a(\delta, \delta_\xi)$ is identically zero as $\xi^a$ is a Killing vector field. Hence there is trivially no leakage of the symplectic structure from the universal horizon. Therefore, eq.~\eqref{eq:hamiltonian definition} still makes sense, i.e., it does not depend upon which slice is used for the integration. $\Sigma$ is now any slice extending from the spatial infinity $i_o$ to the universal horizon and intersecting the universal horizon at a spacelike cross-section $\mathcal{H}$. \par
In order to evaluate the variation $\delta Q_\xi$ and the contribution of $\theta_\xi^{[a}\xi^{b]}$ to the variation of Hamiltonian in eq.~\eqref{eq:hamiltonian_wald} we need to specify the boundary conditions that the fields and their variations must obey.
The asymptotically flat boundary conditions at spatial infinity were discussed in refs. \cite{Eling:2005zq, Foster:2005fr} which we presently review. In some cartesian chart the following fall-off conditions are imposed on the fields: 
 $g_{ab} -\eta_{ab} \sim {1}/{r}$,  
 $ \partial g_{ab} \sim {1}/{r^2}$, 
$u^a - t^a \sim {1}/{r}$, 
$\partial u^a \sim{1}/{r^2} $, 
$\delta u^a \sim {1}/{r}$.
For $\xi^a$ which is asymptotically time translation $t^a$ the value of Hamiltonian is the energy
\begin{align}
\mathcal{E}=H_t = \int_{\partial \Sigma} \exd \sigma_{ab}\,\left( Q_t^{ab} -\,B^{[a}t^{b]} \right).
\end{align}
Using eqs.~\eqref{eq:theta_wald} and \eqref{eq:Noether expression for ae} we get a contribution to $\mathcal{E}$ from the Einstein-Hilbert action which is just the ADM energy and a contribution from the {\ae}ther, $\mathcal{E}=\mathcal{E_{EH}}+\mathcal{E_{\ae}}$, where
\begin{subequations}
\begin{align}
\mathcal{E}_{EH} &= \dfrac{1}{16 \pi G_{\ae}} \int \exd^2 x \, r^i (\partial_i g_{jj} - \partial_j g_{ij}), \\
\mathcal{E}_{\ae} &= \dfrac{c_{14}}{8 \pi G_{\ae}} \int \exd^2 x \, (\partial_t u^r + \partial_r u^t).
\end{align}
\end{subequations} 
%For one-parameter family of static, spherically symmetric solutions labelled by the parameter $r_0$, we get for the total energy 
%\begin{align}
%\mathcal{E} = \frac{r_0}{2G_{\ae}} \left(1-\dfrac{c_{14}}{2}\right).
%\end{align}.
\par
Next, we discuss the boundary conditions on the universal horizon. We refer the reader to the appendix for the definition of the quantities appearing in the following equations. We require that on the universal horizon $\delta u^a = 0$ and $\delta g_{ab} = \delta \gamma_{ab}$, where $\gamma_{ab}$ is the induced metric on the cross-section of the universal horizon orthogonal to the spatial vector $s^a$, i.e., $\gamma_{ab} = g_{ab} + u_a u_b - s_a s_b$. In particular, $\delta s^a$ is also equal to zero on the universal horizon. In effect, we require that a spherically symmetric black hole is perturbed in such a way that the only change is in the expansion of the two-sphere cross-section of the universal horizon orthogonal to the Killing vector $\xi^a$ (and hence $s^a$). 
\par
We are now ready to evaluate the Noether charge in eq.~\eqref{eq:Noether expression for ae} on the universal horizon.  Kinematical consequences of spherical symmetry and staticity are used throughout and are discussed in the appendix.  
Using for the integration on the cross-sections of the universal horizon  $d\sigma_{ab} = -2u_{[a}s_{b]} \, \exd^2x$, a straight-forward calculation gives
\begin{align}
Q_\xi =\frac{1}{8 \pi G_{\ae}} \oint \exd^2x \sqrt{\gamma} \left[\kappa_{UH}(1-c_{13})-c_2\|\xi\|_{UH} K_{UH}\right].
\end{align}
Now owing to the staticity and spherical symmetry of the solutions, the integrand in the Noether charge is constant on the whole universal horizon and can be taken out of the integral. In ref.~\cite{Berglund:2012bu} it was found that asymptotically flat, static, spherically symmetric black hole solutions of Einstein-{\AE}ther theory form a one parameter family. If we further restrict the phase space to allow only one-parameter family of black-hole solutions, then dimensional analysis can be used to shift the variations as 
\begin{subequations}
\label{variation shift}
\begin{align} 
\sqrt{\gamma}\,2 \delta \kappa_{UH} = - \delta(\sqrt{\gamma}) \kappa_{UH}, \\
\sqrt{\gamma}\,2 \delta K_{UH} = - \delta(\sqrt{\gamma}) K_{UH}. 
\end{align}
\end{subequations}
and we get for the variation of $Q_\xi$,
\begin{align}
\label{eq:Noether charge_UH}
\delta Q_\xi = \frac{1}{16 \pi G_{\ae}} \left[\kappa_{UH}(1-c_{13})-c_2\|\xi\|_{UH} K_{UH}\right] \delta (\oint \exd^2x \sqrt{\gamma}).
\end{align}
Next we calculate the $\int \exd \sigma_{ab}\theta^{[a}_\xi \xi^{b]}$ term in eq.~\eqref{eq:hamiltonian_wald}. There is a contribution to this from the terms which come solely from the Einstein-Hilbert Lagrangian (the first line in eq.~\eqref{eq:theta_wald}) and there is a contribution from the {\ae}ther-dependent terms (the second line in eq.~\eqref{eq:theta_wald}). The only contribution from the  {\ae}ther-dependent terms is 
\begin{align}
\label{eq:theta_aether}
-\frac{1}{16 \pi G_{\ae}}  \oint\exd^2x\sqrt{\gamma}\, Y^{(ab)}u^c \delta g_{ab} u_c \|\xi\|_{UH} = -\frac{1}{8 \pi G_{\ae}} \oint \delta(\exd^2x\sqrt{\gamma}) \left(c_{13} \hat{K}_{UH} + 2c_2  K_{UH}\right) \|\xi\|_{UH},
\end{align}
while the Einstein-Hilbert terms give
\begin{subequations}
\label{eq:theta_EH}
\begin{align}
\frac{1}{16 \pi G_{\ae}}\oint \exd^2x \sqrt{\gamma}\, \|\xi\|_{UH} u^a \nabla^b \delta g_{ab} &= -\frac{1}{16 \pi G_{\ae}}\oint \delta(\exd^2x\sqrt{\gamma}) \|\xi\|_{UH} \hat{K}_{UH} ,  \\
\frac{1}{16 \pi G_{\ae}}\oint \exd^2x \sqrt{\gamma}\, \|\xi\|_{UH} g^{ab} u^c \nabla_c \delta g_{ab} &= \frac{1}{16 \pi G_{\ae}}\oint \exd^2x \sqrt{\gamma}\, \|\xi\|_{UH}\,2\,\delta K_{UH}.
\end{align}
\end{subequations}
Again, restricting the phase space to consist of only one parameter family of static and spherically symmetric solutions, using eqs.~\eqref{variation shift} and combining the contributions from eqs.~\eqref{eq:theta_aether} and \eqref{eq:theta_EH} we get
\begin{align}
\label{eq:theta_UH}
\oint \exd \sigma_{ab}\,\theta^{[a}\xi^{b]} = -\frac{1}{16 \pi G_{\ae}}\oint \delta \left( \exd^2x \sqrt{\gamma}\right)\,
\left( \kappa_{UH} + c_{13} K_{UH}\|\xi\|_{UH} - c_{13} \kappa_{UH} + 2 c_2 K_{UH}\|\xi\|_{UH}\right)
\end{align}
Finally, from eqs.~\eqref{eq:Noether charge_UH} and \eqref{eq:theta_UH} we get the contribution to $\delta H_\xi$ from the universal horizon,
\begin{align}
\label{eq:delta hamiltonian_UH}
\delta H_\xi \Big| _{UH} = \frac{1}{8 \pi G_{\ae}}\left[\kappa_{UH}(1-c_{13}) + \frac{c_{123}}{2} K_{UH} \|\xi\|_{UH} \right] \delta (\oint \exd^2x \sqrt{\gamma}).
\end{align}
Since the contribution from infinity $\delta H_\xi \Big| _{\infty}$ is identified as the change in energy $\delta \mathcal{E}$ of the black hole, we have a thermodynamical first-law form for the mechanics of black holes in Einstein-{\AE}ther theory,
\begin{align}
\delta \mathcal{E} = T \delta S,
\end{align}
where in analogy with general relativity, we have called $S=A/4$ the entropy of the universal horizon and $T=\dfrac{1}{2 \pi G_{\ae}}\left[\kappa_{UH}(1-c_{13}) + \dfrac{c_{123}}{2} K_{UH} \|\xi\|_{UH} \right]$ the temperature of the universal horizon. Our expression matches with the one found in ref.~\cite{Berglund:2012bu}. If we further invoke the result of ref.~\cite{Berglund:2012fk} that what we have called temperature is really the temperature at which the universal horizon radiates then this is a strong hint at the possibility that the universal horizons carry entropy too.
\section{Discussion}
\label{sec:Discussion}
In this paper we derived the first-law for universal horizons in the Einstein-{\AE}ther theory using the Noether charge formalism of Wald. Owing to the divergence of the {\ae}ther on the bifurcation surface, the Noether-charge method is not directly applicable for the Killing horizon in Einstein-{\AE}ther theory. However, we saw that for the universal horizons one can adapt the Noether-charge formalism and obtain a first law. In doing so however, we restricted the phase space to consist of the asymptotically flat, static, spherically symmetric solutions of the theory. These solutions were shown to form  a one-parameter family in ref.~\cite{Berglund:2012bu}. In our treatment we had to use this fact to use the dimensional arguments in shifting the variational derivatives as in eq.~\eqref{variation shift}. It would be useful to find another way such that one does not have to use this fact, one could then prove the first-law just like in general relativity for arbitrary perturbations which are not necessarily stationary. It would also be of interest to see if one could prove a physical process version of the first-law. \par
Number of questions remain unanswered though. What is the nature of signals originating close to the universal horizons? Do they show a peeling-off property as seen for the light rays close to the Killing horizons. What would be a useful definition of surface gravity for universal horizons? This question is partially studied in ref.~\cite{Cropp:2013zxi}. It would be interesting to see if there is a relation between some notion of surface gravity and the temperature of the universal horizon. Although, the temperature occurring in the first-law agrees with the temperature calculated using the tunnelling formalism in ref.~\cite{Berglund:2012fk}, it does not tell us what the asymptotic observer sees. These issues are currently under study and will be reported elsewhere \cite{Bethan}. It also remains to be seen if the universal horizon serves to save the GSL in Lorentz violating theories in the way discussed in the introduction. Finally, stability of universal horizons and their existence in the axisymmetric case is an important and interesting question.\par

\appendix*
\section{Kinematical consequences of staticity and spherical symmetry}
\label{appendix}
We collect here useful identities which result from symmetry arguments. We closely follow the ref.~\cite{Berglund:2012bu}, the reader is referred to that paper for a more detailed discussion. \par
For spherically symmetric solutions, the {\ae}ther must be hypersurface orthogonal \cite{Eling:2006ec}. Denote the hypersurfaces orthogonal to $u^a$ as $\Sigma_u$. By spherical symmetry we can foliate $\Sigma_u$ by two-spheres $S_u$. We now have a preferred basis of frame defined at every point in spacetime. It consists of a unit timelike vector which is simply the {\ae}ther $u^a$, a unit spacelike vector denoted by $s^a$ which is orthogonal to $S_u$ and contained in $\Sigma_u$ (and hence is orthogonal to $u^a$), and two more spacelike vectors which span the $S_u$ and which we will not need here. \par 
Again due to spherical symmetry, any vector can be written in terms of its components along $u^a$ and $s^a$. For example, the Killing vector $\xi^a$ can be written as
\begin{align}
\label{xi decomposition}
\xi^a = -(\xi \cdot u) u^a + (\xi \cdot s) s^a. 
\end{align}
Similarly, any rank-two tensor can be decomposed into its components along $u_a u_b, s_a s_b, u_{[a}s_{b]}$ and $u_{(a}s_{b)}$. For example, the extrinsic curvature of $\Sigma_u$ hypersurface can be decomposed as 
\begin{align}
\label{Kab decomposition}
K_{ab} = K_0 s_a s_b + \frac{\hat{K}}{2} \gamma_{ab},
\end{align}
where $\gamma_{ab}$ is the induced metric on the two-spheres $S_u$. Trace of the extrinsic curvature is given by $K=K_0 + \hat{K}$. The derivative of the Killing vector can be decomposed as
\begin{align}
\label{xi derivative}
\nabla_a \xi_b = -2\kappa u_{[a}s_{b]},
\end{align}
where $\kappa = \sqrt{- \dfrac{1}{2}(\nabla_a \xi_b)(\nabla^a \xi^b)}$. Derivatives of $u^a$ and $s^a$ can be decomposed as 
\subequations
\label{u and s decomposition}
\begin{align}
\nabla_a u_b &= -(a \cdot s) u_a s_b + K_{ab} \\
\nabla_a s_b &= K_0 s_a u_b + K^{(s)}_{ab},
\end{align}
where $K^{(s)}_{ab} = -(a \cdot s) u_a u_b + \dfrac{\hat{k}}{2} \gamma_{ab}$, is the extrinsic curvature of $\Sigma_s$ hypersurfaces. $\hat{k}$ is the trace of extrinsic curvature of the two spheres $S_u$ as embedded in $\Sigma_u$ while $\hat{K}$ in eq.~\eqref{Kab decomposition} is the trace  of extrinsic curvature of the two spheres $S_u$ as embedded in $\Sigma_s$ hypersurfaces. \par
Using the decompositions in eqs.~\eqref{u and s decomposition}, the ''surface gravity" $\kappa$ can now be calculated as
\begin{align}
\kappa = -(a \cdot s)(u \cdot \xi) + K_0 (s \cdot \xi),
\end{align}
and its value on the universal horizon is $\kappa_{UH} = K_{0,UH} \| \xi \|$ since on the universal horizon $u \cdot \xi = 0$ and hence $s \cdot \xi = \| \xi\|$.

%Hypersurface orthogonality of the {\ae}ther implies that its twist must vanish \cite{Wald:1984rg}, i.e., $u_{a[}\nabla_b u_{c]} = 0$. Similarly, spherical symmetry and staticity also implies that $s^a$ is orthogonal to hypersurfaces which we denote by $\Sigma_s$ and thus $s_{a[}\nabla_b s_{c]} = 0$. Taking the components of these equations along $u^a$ and $s^a$ we get
%\subequations
%\begin{align}
%\nabla_{[a}u_{b]} &= -(a\cdot s) u_{[a}s_{b]}  \\
%\nabla_{[a}s_{b]} &= -K_0 u_{[a}s_{b]}
%\end{align}

\begin{acknowledgments}
I thank Stefano Liberati and Thomas Sotiriou for stimulating discussions. I also thank Abhay Ashtekar and Ted Jacobson for comments and discussion. 
\end{acknowledgments}

\bibliography{ThermodynamicsOfUnivHorizonsRefs}

\end{document}